\documentstyle[epsf]{mn}
\newif\ifAMStwofonts

\newcommand{\be}{\begin{equation}}
\newcommand{\ee}{\end{equation}}
\newcommand{\bgo}{\begin{eqnarray}}
\newcommand{\ego}{\end{eqnarray}}
\newcommand{\bg}{\begin{eqnarray}}
\newcommand{\eg}{\end{eqnarray}}
\newcommand{\ba}{\begin{eqnarray}}
\newcommand{\ea}{\end{eqnarray}}
\newcommand{\brr}{\begin{array}}
\newcommand{\err}{\end{array}}
\newcommand{\bc}{\begin{centre}}
\newcommand{\ec}{\end{centre}}

\newcommand{\h}{$h^{-1}$\thinspace}
\ifoldfss
  \ifCUPmtlplainloaded \else
    \NewTextAlphabet{textbfit} {cmbxti10} {}
    \NewTextAlphabet{textbfss} {cmssbx10} {}
    \NewMathAlphabet{mathbfit} {cmbxti10} {} 
    \NewMathAlphabet{mathbfss} {cmssbx10} {} 
  \fi
  \ifAMStwofonts
    \ifCUPmtlplainloaded \else
      \NewSymbolFont{upmath} {eurm10}
      \NewSymbolFont{AMSa} {msam10}
      \NewMathSymbol{\upi}     {0}{upmath}{19}
      \NewMathSymbol{\umu}     {0}{upmath}{16}
      \NewMathSymbol{\upartial}{0}{upmath}{40}
      \NewMathSymbol{\leqslant}{3}{AMSa}{36}
      \NewMathSymbol{\geqslant}{3}{AMSa}{3E}

       \let\le=\leqslant
       \let\ge=\geqslant
    \fi
  \fi
\fi 

\ifnfssone
  \newmathalphabet{\mathit}
  \addtoversion{normal}{\mathit}{cmr}{m}{it}
  \addtoversion{bold}{\mathit}{cmr}{bx}{it}
  \newmathalphabet{\mathbfit} 
  \addtoversion{normal}{\mathbfit}{cmr}{bx}{it}
  \addtoversion{bold}{\mathbfit}{cmr}{bx}{it}
  \newmathalphabet{\mathbfss} 
  \addtoversion{normal}{\mathbfss}{cmss}{bx}{n}
  \addtoversion{bold}{\mathbfss}{cmss}{bx}{n}
  \ifAMStwofonts
    \ifCUPmtlplainloaded \else
      \UseAMStwoboldmath
      \makeatletter
      \new@mathgroup\upmath@group
      \define@mathgroup\mv@normal\upmath@group{eur}{m}{n}
      \define@mathgroup\mv@bold\upmath@group{eur}{b}{n}
      \edef\UPM{\hexnumber\upmath@group}
      \new@mathgroup\amsa@group
      \define@mathgroup\mv@normal\amsa@group{msa}{m}{n}
      \define@mathgroup\mv@bold\amsa@group{msa}{m}{n}
      \edef\AMSa{\hexnumber\amsa@group}
      \makeatother
      \mathchardef\upi="0\UPM19
      \mathchardef\umu="0\UPM16
      \mathchardef\upartial="0\UPM40
      \mathchardef\leqslant="3\AMSa36
      \mathchardef\geqslant="3\AMSa3E

       \let\le=\leqslant
       \let\ge=\geqslant
    \fi
  \fi
\fi 

\ifnfsstwo
  \DeclareMathAlphabet{\mathbfit}{OT1}{cmr}{bx}{it}
  \SetMathAlphabet\mathbfit{bold}{OT1}{cmr}{bx}{it}
  \DeclareMathAlphabet{\mathbfss}{OT1}{cmss}{bx}{n}
  \SetMathAlphabet\mathbfss{bold}{OT1}{cmss}{bx}{n}
  \ifAMStwofonts
    \ifCUPmtlplainloaded \else
      \DeclareSymbolFont{UPM}{U}{eur}{m}{n}
      \SetSymbolFont{UPM}{bold}{U}{eur}{b}{n}
      \DeclareSymbolFont{AMSa}{U}{msa}{m}{n}
      \DeclareMathSymbol{\upi}{0}{UPM}{"19}
      \DeclareMathSymbol{\umu}{0}{UPM}{"16}
      \DeclareMathSymbol{\upartial}{0}{UPM}{"40}
      \DeclareMathSymbol{\leqslant}{3}{AMSa}{"36}
      \DeclareMathSymbol{\geqslant}{3}{AMSa}{"3E}

       \let\le=\leqslant
       \let\ge=\geqslant
    \fi
  \fi
\fi 

\ifCUPmtlplainloaded \else
  \ifAMStwofonts \else 
    \def\upi{\pi}
    \def\umu{\mu}
    \def\upartial{\partial}
  \fi
\fi

\title[Searching for the scale of homogeneity]
{Searching for the scale of homogeneity}
\author[V.J. Mart\'{\i}nez et al.]
{Vicent J. Mart\'{\i}nez,$^{1}$\thanks{Email: martinez@quasar.daa.uv.es}
Mar\'{\i}a--Jes\'us 
Pons--Border\'{\i}a,$^{2}$\thanks{Email: pons@astro1.ft.uam.es}
Rana A. Moyeed$^{3}$\thanks{Email: rmoyeed@plymouth.ac.uk}\cr and
Matthew J. Graham$^{1,4}$\thanks{Email: m.j.graham@uclan.ac.uk} \\
$^1$Departament d'Astronomia i Astrof\'{\i}sica,
Universitat de Val\`encia, E--46100 Burjassot, Val\`encia,
Spain \\
$^2$Departamento de F\'{\i}sica Te\'orica, Universidad Aut\'onoma de
Madrid,
28049 Cantoblanco, Madrid, Spain \\
$^3$School of Mathematics and Statistics,
University of Plymouth, Drake Circus, Plymouth PL4 8AA, U.K. \\
$^4$Centre for Astrophysics, University of Central Lancashire, Preston 
PR1 2HE, U.K.
}
 
\date{Accepted 1997 ???? ??.
Received 1997 ???? ??;
in original form 1996 ???? ??}

\begin{document}
 
\maketitle
\begin{abstract}

We introduce a statistical quantity, known as the $K$ function, related
to the integral of the two--point correlation function.
It gives us straightforward information about the scale where
clustering dominates and the scale at which homogeneity
is reached. We evaluate the correlation dimension, $D_2$, as the
local slope of the log--log plot of the $K$ function.
We apply this statistic
to several stochastic point fields, to three numerical simulations
describing the distribution of clusters and finally to real galaxy
redshift surveys.
Four different galaxy catalogues have been analysed using this
technique: the Center for Astrophysics I, the Perseus--Pisces 
redshift surveys (these two lying in our local neighbourhood),
the Stromlo--APM and the 1.2 Jy {\it IRAS} redshift surveys 
(these two encompassing a larger volume).
In all cases, this cumulant
quantity shows the fingerprint of the transition to homogeneity.
The reliability of the estimates is clearly demonstrated by
the results from controllable point sets, such as the segment Cox
processes.
In the cluster distribution models, as well as in the real galaxy catalogues,
we never see long plateaus when plotting $D_2$ as a function of the
scale, leaving no hope for unbounded fractal distributions.

\end{abstract}
 
\begin{keywords}
methods: statistical; galaxies: clustering;
large--scale structure of Universe
\end{keywords}
 
\section{Introduction}
The standard cosmology is based on the assumption that the Universe
must be homogeneous on very large scales. Several pieces of evidence support 
this assumption: the homogeneity and isotropy of the microwave background
radiation \cite{cobe} and some aspects of the large scale distribution
of matter \cite{peb89} seem to strongly advocate uniformity on
scales bigger than about $200 \, h^{-1}$ Mpc 
(where $H_0= 100 h$ km s$^{-1}$ Mpc$^{-1}$).

However the presence of very large features in the galaxy distribution
like
the Bootes void \cite{kir81} or the Great Wall \cite{gel89}
which span a scale length of the
order of $100 \, h^{-1}$ Mpc calls the actual scale of homogeneity
into question. Moreover other authors consider the assumption of
homogeneity just a theoretical prejudice not necessarily supported
by the observational evidence quoted above. They defend the alternative
idea of an unbounded fractal cosmology \cite{col92}. 
Guzzo (1997) argues against this interpretation on the basis of a
careful handling of the data.
 
The spatial two--point correlation function is the statistical tool mainly
used to describe the clustering in the Universe (Peebles 1980, 1993). 
However, because of the 
integral constraint \cite{p80}, one cannot estimate it at very large distances
from the currently available redshift surveys.
In order to study 
clustering in the regime where it is not very strong, we have only two 
possibilities: either we extend the size of the redshift catalogues or
we use  alternative statistical descriptors. The approach described in this
paper points in the latter direction. 
In the same line, other authors \cite{fis93,par94,tad96} have
tried to measure the power--spectrum on large scales directly from galaxy 
catalogues. 
Einasto \& Gramman (1993) studied the transition to homogeneity by means of 
the power--spectrum and found a relation between the correlation transition 
scale and the spectral transition scale (turnover in $P(k)$).  

We introduce the quantity called $K(r)$, which is related to the
correlation function $\xi(r)$. The novelty of our approach lies essentially 
in the fact that we shall use a cumulant quantity instead of
a  differential quantity such as $\xi(r)$. 
Although for a point process the functions
$\xi(r)$ and $K(r)$ are well defined, what we measure from the
galaxy catalogues are just estimators of those functions.
One of our main claims is that the estimators for $K(r)$ are 
more reliable than the most currently used estimators for $\xi(r)$ and
that makes its use recommendable (especially in three-dimensional processes
and at large scales) despite its somewhat less informative character.

\section{The $K$ function and the correlation dimension}

Within the field of the statistical analysis of point fields,
new techniques and estimators of the clustering of spatial
point patterns have been developed in the past decades.
Unfortunately, the connection between this set of
scholars and cosmologists is not 
as important today as it was in the late fifties
when the Berkeley statisticians Neymann and Scott carried out
an intensive programme about the analysis of the Lick catalogue
(Neymann \& Scott 1952, 1955).
Today, one of the most popular summary statistics for point patterns is
the $K$--function (Bartlett 1964, Ripley 1976, 1977, 1981).
Let us introduce this quantity using the
terminology employed by statisticians and stressing its connections with
the quantities used in cosmology (see also Szapudi \& Szalay 1997 for
a different application of Ripley's statistic to cosmology).

Let us consider a point process acting on a region $D \subset \bmath{R}^3$ 
with volume $V$
whose output is a collection of positions of $N$ galaxies
(or clusters of galaxies) \{${\bf x}_i$\}. If we take
two infinitesimal volumes $dV_1$ and $dV_2$ around ${\bf x}_1$ and
${\bf x}_2$ respectively, the joint probability of there being a point 
lying in each of these volumes reads:
\begin{equation}
dP = \lambda_2 ({\bf x}_1, {\bf x}_2) dV_1 dV_2,
\end{equation}
where $\lambda_2$ is the so-called second order intensity function
\cite{dig83}.
If the process is stationary (invariant under translation) and isotropic
(invariant under rotation), then
$\lambda_2({\bf x}_1, {\bf x}_2) = \lambda_2(|{\bf x}_1- {\bf x}_2|)$.
The two-point correlation function can be expressed by means
of it as:
\begin{equation}
\xi(r)= \frac{\lambda_2(r)}{n^2} - 1
\end{equation}
where $r=|{\bf x}_1-{\bf x}_2|$ and $n$ is the mean number density in a
fair sample.

The second--order cumulative function $K(r)$ is defined so that 
the expected number of neighbours a given galaxy will have at a distance
less than $r$ is $n K(r)$. Therefore its relation with the two--point 
correlation function is 

\begin{equation}
K(r) = \int_0^r 4\pi  s^2 (1+\xi(s)) ds 
\end{equation}
For a homogeneous Poisson process this function is just

\begin{equation}
K_{\rm \tiny Pois} (r) = {4 \pi \over 3} r^3. \label{kpois}
\end{equation}

\subsection{Relation with other cumulant quantities}
Other second--order cumulant functions have been used in the statistical
analysis of the large scale structure in the Universe. Within the context 
of the scaling or multifractal approach the partition function $Z(q,r)$ 
introduced in the description of the galaxy clustering by Mart\'{\i}nez
et al. (1990) is formally related with $K(r)$ for the second moment
$(q=2)$ by $Z(2,r)=K(r)/V$ where $V$ is the volume of the sample.
Borgani et al. (1994) perform an exhaustive 
analysis of the cluster distribution
in both real catalogues of clusters of galaxies and simulations by means
of the $Z(q,r)$ function. In that paper the authors give an expression
for using this partition function when a selection function has to be
considered.
The dependence of the results of $Z(q,r)$ of the particular 
volume of the sample
has led some authors (Dom\'{\i}nguez--Tenreiro, G\'omez--Flechoso and 
Mart\'{\i}nez 1994) to normalize $Z(2,r)$ in order to get a function
$Z_{\rm \tiny norm}$  which coincides for different 
samples (within different volumes)
in the homogeneous regime. Basically this normalization makes 
$Z_{\rm  \tiny norm}$ equivalent to $K(r)$.

Peebles (1980) introduced the moments of the counts of neighbours
$\langle N \rangle_r$ as the mean count of objects in balls of radius
$r$ excluding the central one. By definition  $\langle N \rangle_r$
is the correlation integral $C(r)$ used by Mart\'{\i}nez et al. (1995)
in the analysis of the multiscaling properties of the matter
distribution and is related with $K(r)$ simply by 
$\langle N \rangle_r = n K(r)=C(r)$.
In fact, other authors (Mart\'{\i}nez \& Coles 1994; McCauley 1997)
have chosen the normalization for the partition
function $Z(q,r)$ in such a way that $\langle N \rangle_r = Z(2,r)$
directly.

Taking the shape of the $K$ function for a random
object distribution (Poisson process) into account, one can just consider the 
difference between $K(r)$ and $K_{\rm \tiny Pois}(r)$ which leads to 
another
cumulant quantity commonly used in the statistical description of the
galaxy clustering, $4 \pi J_3(r) = K(r) -K_{\rm \tiny Pois}(r)$, where
(Peebles 1980, 1993)
\begin{equation}
J_3(r) = \int_0^r \xi(s) s^2 ds.
\end{equation}

Finally, if one considers the quotient instead of the difference, one
gets the integral quantity used by Coleman \& Pietronero (1992):

\begin{equation}
\Gamma^{\ast}(r) = {n K(r) \over K_{\rm \tiny Pois}(r)} ,
\end{equation} 
(see also Cappi et al. (1998) for a discussion on the methods used
by Pietronero and collaborators).

The advantages of the use of $K(r)$ with respect to other cumulant quantities
are the following:
\begin{enumerate}
\item
$K(r)$ is well normalized. We can compare directly the $K$ function of
different samples, with different number density and within different
volumes without extra normalization.
\item
$K(r)$ is a well--known quantity in the field of spatial statistics and
several analytical results regarding its shape and variance are already
available for a variety of point processes.
\item
It is very important to estimate the quantity $K(r)$ directly from
the data and not through numerical integration of $1+\xi(r)$, which 
introduces artificial smoothing of the results. 
Several edge--corrected unbiased estimators are available for $K(r)$.
In the context of the present application, the most appreciable
properties an estimator must hold are to have little variance and not
to introduce spurious homogeneity by means of the edge--correction.
In the next subsection we comment on different estimators for $K(r)$.
\end{enumerate}
\subsection{Estimators}

We shall make the assumption that the process under consideration
is stationary and isotropic.
Such a process is also referred to as
`statistically homogeneous' \cite{p93}. Note, however, that when in
this paper we talk about `the scale of homogeneity', we mean the scale
at which the spatial distribution of the objects is uniform or
indistinguishable from a homogeneous Poisson process. 
Nevertheless, some cluster processes are still stationary and isotropic.

There exist several estimators for $K$. 
A comparison of some of them can be found in Doguwa \& Upton (1989).
From the definition of $K$ and ignoring the edge effects one could
consider the following naive estimator
\begin{equation}
\hat{K}_{\rm N}(r) =
{V\over N^2} \sum_{i=1}^N \sum^N_{\scriptstyle
j=1 \atop\scriptstyle j\ne i }
{\theta(r-|{\bf x}_i -{\bf x}_j|)}, \label{naive}
\end{equation}
where $\theta$ is Heaviside's step function, whose value is 1 when the
argument is positive and 0 otherwise.
Obviously for a finite sample this estimator will provide values
for $K$ smaller than the true values since neighbours outside the boundaries
are not considered.
One possible solution is to consider only points in an inner region as
centres of the balls for counting neighbours. The points lying in the
outer region, a buffer zone (Upton \& Fingleton 1987; Buchert \& Mart\'{\i}nez 
1993), take part in the
estimator just as points which could be seen as neighbours at a given distance
$r$ of the points in the inner region. 
The inner region might shrink as $r$ increases. However, this solution 
leads to biases (the sample is not uniformly selected),
wastes a lot of data, and obviously increases
the variance of the estimator \cite{dog89}.
The standard solution adopted in the statistical studies of the large--scale
structures is to account for the unseen neighbours outside the sample window
by means of the following edge--corrected estimator
\begin{equation}
\hat{K}_{\rm DU}(r) =
{V\over N^2} \sum_{i=1}^N \sum^N_{\scriptstyle
j=1 \atop\scriptstyle j\ne i }
{\theta(r-|{\bf x}_i -{\bf x}_j|)\over f_i(r)}, \label{estim2}
\end{equation}
where $f_i(r)$ is the fraction of the volume of the sphere of radius $r$
centred on the object $i$ which falls within the boundaries of the sample.
This kind of edge--correction has been used by Borgani et al. (1994) when
calculating the partition function $Z(q,r)$ used in the multifractal analysis.
In the field of spatial statistics it had been introduced by Doguwa and Upton 
(1989). Although this estimator has good properties, it could be
slightly biased \cite{sto94}.
The most commonly used unbiased edge--corrected estimator in the analysis
of point processes
is Ripley's estimator, which under our hypotheses reads \cite{rana}:
\begin{equation}
\hat{K}_{\rm R}(r) =
{V\over N^2} \sum_{i=1}^N \sum^N_{\scriptstyle
j=1 \atop\scriptstyle j\ne i }
{\theta(r-|{\bf x}_i -{\bf x}_j|)\over \omega_{ij}}, \label{estim}
\end{equation}
where the weight $\omega_{ij}$ is an edge correction equal to the
proportion
of the {\sl area} of the sphere centred at ${\bf x}_i$ and passing
through
${\bf x}_j$ that is contained in $D$; in other words, $\omega_{ij}$ is
the conditional probability that the $j$th point is observed given that it
is at a distance
$r$ from the $i$th point. This correction is suitable for stationary and 
isotropic processes and is illustrated in Fig.~\ref{edge}. The unweighted 
$K$ function will be negatively
biased because we do not observe events outside the sampling window, so the
observed counts from events which are less than a distance $r$ from the
boundary will be artificially low.

\begin{figure}
\vspace{1.cm}
\caption{An illustration of the weights used in the estimator of $K$
(equation 9) in 2 dimensions. The rectangle represents the boundary of 
the sample.
In this case, $w_{ij}$ is the proportion of the circumference of
the circle centered at ${\bf x}_i$, passing through ${\bf x}_j$, lying
within the boundary of the sample.
Depending on the relative positions of the galaxies with respect to the
boundary, different cases are illustrated: (a) $w_{ij}=w_{ji}=1$;
(b) $w_{ij}=1$, $w_{ji} < 1$; (c) $w_{ij} < 1$, $w_{ji} < 1$. 
It is clear from the plot that we weight the observed neighbour
${\bf x}_j$ of the galaxy ${\bf x}_i$ lying at a distance $r$ (the radius
of the circle) from it by the inverse of the probability that such 
a neighbour would be observed.}
\label{edge}
\end{figure}

It is still possible that the best estimator depends of the kind of point 
process to be studied (clustered or regular) and even on the particular 
scale range, (see Stoyan \& Stoyan (1998) for a discussion on improved 
estimators). We have chosen
the estimator $\hat{K}_{\rm R}(r)$ because of its well known good performance
in a variety of cases.

In Baddeley et al. (1993) an analytic expression
for $\omega_{ij}$ is given in the case $D$ is a cube.
In order to ensure that the border correction is as free 
of error as possible, we have chosen to generate the synthetic
samples we want to analyse in a region shaped in this way.
Note however that 
we introduce a certain bias when we estimate $n$ through $N/V$ but we are,
on the other hand, making full use of all sample points.
For all the mentioned estimators it is possible to build the corresponding
versions for flux--limited samples by simply adding a weighting factor 
representing the selection function.

\subsection{The correlation dimension}

If scaling of the first moment of the count of neighbours holds, then
$C(r)$ is proportional to $r^{D_2}$ and $K(r)$ as well, since $n$ is
constant when looking at the whole region. The exponent $D_2$ is
known as the correlation dimension and it tends to 3 on the scale
$r$ at which homogeneity is reached, hence its importance.

Once we compute $K$, we obtain the local dimension $D_2$ as the slope of
a five-point log-log linear regression
on the function $K(r)$ (as in Borgani et al. 1994). This local
correlation dimension can be considered as a {\it sliding window}
estimate (through the scale) of the fractal dimension \cite{dub89}.
Any long plateau
for a significant range of scales will be the fingerprint of a fractal
range.
The tendency to homogeneity will be described by an increasing behaviour
of the local dimension $D_2$ as a function of the scale, towards the value
consistent with homogeneity $D_2 =3$.
This test is much stronger than just to fit a straight line to a log--log plot,
which could lead to wrongly interpret as fractals sets which clearly are 
not, as we show in Section 5.3 (see also Stoyan 1994, McCauley 1997).

\section{The $K$ function on stochastic point processes}

In order to test the usefulness of the statistic $K$ and the
accuracy of its estimator [equation (\ref{estim})] we have calculated
$K$ and $D_2$ for three different point samples.
They are controllable point processes for which we have some {\it a
priori}
knowledge of the expected behaviour of the $K$ function.
In order to compute as well the deviations from this theoretical
behaviour
when the empirical value is obtained on a point sample, we have analysed 
several realizations of each process, simulated in a cube of side
300~$h^{-1}$ 
Mpc. Let us briefly describe the models analysed here.

\begin{figure}
\vspace{1.cm}
\caption{One of the realizations of each of the stochastic point
processes
analysed here: Poisson, Soneira--Peebles and Voronoi
models described in Section 3. The side length is 300 $h^{-1}$ Mpc.}
\label{cubes}
\end{figure}

i) Poisson sample. We generate 10 of these samples containing
1000 points each in order to test if our quantities behave in the
expected way in the absence of any pattern.

ii) Soneira--Peebles model (S-P).  Starting from a sphere of radius $R$
and
following Soneira \& Peebles (1978), we randomly place  $\eta$ spheres
of
radius
$R/\lambda$.  In each of the resulting smaller spheres we do the same
with
spheres of radius $R/{\lambda^2}$. We repeat this $L$ times and consider
the centres of the $\eta^L$ spheres in the last level as our study
objects. Although Soneira \& Peebles (1978) add up $N_r$ realizations 
of this kind, we simply generate
one, which is known to be a single fractal \cite{mar90}. The model is 
included inside a cube of side 
$300~h^{-1}$ Mpc. The values of the
parameters we have
chosen are $\eta=2$, $\lambda=2$, $L=10$, and we have not allowed
spheres in the same level to overlap. The expected value of $D_2$ can be
analytically
calculated for this model through the relation $D_2=(\log \eta)/(\log
\lambda)$ and, for this choice of the parameters, it turns out to be $D_2=1$.
We have generated 10 of such samples.

iii) Voronoi sample. Essentially it is built \cite{rien}
by choosing points at random as centres of the cells of a Voronoi
tessellation,
which is produced by drawing planes equidistant to the nearest centres.
Then a certain amount of objects (events of the point process)
following a Gaussian distribution  are put on and around the filaments
which result of the intersection of two of those planes. We have analysed
5 realizations containing around 2500 points each.

All these point processes are shown in Fig.~\ref{cubes} and
now we shall comment on the results of the $K$--function for the first three 
point processes, which are
graphically represented in Fig.~\ref{kppfig}. 
For each sample, we plot the average
of the function $K(r)$  calculated over the 10 realizations in the
range of scales [10, 100] $h^{-1}$ Mpc, except for the Voronoi sample,
which begins at $r=1~h^{-1}$ Mpc. We plot the corresponding standard
deviation as error bars. These are usually
smaller than the size of the dot, probing thus the stability of the
estimator.
The dotted line represents the theoretical behaviour for a Poisson process.
Let us mention that the depth up to which a statistic can be reliably
calculated depends on the statistic itself and on the estimator used
as well as on the geometry of the region.
In our case, we have calculated $K(r)$ for $r\in [5, 150]\,h^{-1}$ 
Mpc and it follows the same trend as in the shorter interval [10,100]
$h^{-1}$ Mpc
that we have plotted, except for more significant noise at very small
scales. On the top of
each panel we show the local dimension $D_2$ calculated as stated in
Section 2. Note that we do not calculate $D_2$ basing ourselves on the 
$K$ of each individual 
sample but on the average $K$ of all the samples of a same process.
Error bars are now 1 $\sigma$ uncertainties in the five--point weighted
least squares fit.

\begin{figure}
\vspace{1.cm}
\caption{The functions $K(r)$ and $D_2(r)$ for the
point processes i), ii) and iii).
The straight continuous line corresponds to $K_{\rm \tiny Pois}$.
} 
\label{kppfig}
\end{figure}

In the estimation of the $K$ function, we have checked that the lack
of border correction (i.e., if we
take $\omega_{ij}=1 ~\forall i,j$) would underestimate significantly
the amount of clustering, giving a false sense of regularity in the
data.

Let us first point out that $K$ has fully accomplished its mission as a
detector of the scale of homogeneity. For the Poisson distribution we
see that the estimator of $K(r)$ matches quite well the expected behaviour
of equation (\ref{kpois}). 
Moreover the value of the local dimension is nearly
constant for the whole range of scales  and $D_2=3$, with little
fluctuations at small scales due to the small number of points.

The other two models clearly show clustering, since the function
$K(r) \ge K_{\rm \tiny Pois} (r)$.
Note that homogeneity begins to be reached
when the $K$ curve of the sample falls on the theoretical straight line.
The simple S-P clump is a pure fractal and hence it never
reaches homogeneity; on the plot,
this corresponds to the fact that the sample and theoretical lines do
not coincide over the whole range of scales, being both straight lines with
different slopes. The local dimension $D_2$ as a function of the scale $r$
is practically a horizontal line at the theoretically predicted height,
$D_2 \simeq 1$. This plateau is the fingerprint of fractal
behaviour.
The lacunarity of this fractal set, appreciated in Fig. 1, is responsible
for the important error bars in the plot of $D_2$. For larger values of
$r$ the estimates of $K$ and hence of $D_2$ are biased by the
edge correction,
which assumes that the process is stationary and isotropic. Nevertheless
for $r \le 80$ $h^{-1}$ Mpc they give the correct result although the
point process does not fulfill the required assumptions, showing that
the method does {\sl not} introduce spurious homogenization.

The different prescriptions used to generate the analysed point
processes are  reflected in the different shapes
of $K$ and in the behaviour of $D_2$ as a function of the scale.
The Voronoi model used here is based on the population of the filaments;
this can be seen in the fact that, for the smallest scales shown in the
plot ($r < 8$ $h^{-1}$ Mpc), the dimension $D_2 \simeq 1$, corresponding to
filamentary--like structures. At large scales
we can appreciate a continuous variation of the local dimensionality
from $D_2
\simeq 1$ to $D_2 \simeq 3$, corresponding to the scale at which
homogeneity is reached ($r_{\rm \tiny hom} \simeq 50$ $h^{-1}$ Mpc).

\section{The $K$ function on simulations of the cluster distribution}

\begin{figure}
\vspace{1.cm}
\caption{The different $N$--body simulations of the cluster
distribution.
The side of the box is 300 $h^{-1}$ Mpc.}
\label{simulati}
\end{figure}

We have furthermore analysed three sets of 10 $N$--body simulations 
of the distribution of clusters of galaxies
(Croft \& Efstathiou 1994a, 1994b).
They have used a $P^3M$ code to generate these
samples, containing around 1000 clusters each of them in a cube of
comoving side 300 $h^{-1}$ Mpc.

i) Standard CDM. It has been generated from a CDM power spectrum
that applies for models with low baryon density. The chosen values of the
cosmological parameters are $\Omega=1$  and $h=0.5$.

iii) Mixed dark matter (MDM). It contains a massive neutrino component
and CDM in a proportion so that $\Omega_{\rm \tiny total}=1$ and
$\Omega_{\nu}=0.3$, taking $h=0.5$.

iii) CDM with cosmological constant ($\Lambda$--CDM). The values of the
parameters are $\Omega=0.2$, $h=1$ and $\Lambda/(3H_0^2)=0.8$.

All these simulations are shown in Fig.~\ref{simulati}.

\begin{figure}
\vspace{1.cm}
\caption{The standard correlation function $\xi(r)$ for the three
cluster samples. We have plotted the average values over 10 realizations 
(only 8 for MDM) and the error bars correspond to 1 $\sigma$ deviations.}
\label{cf}
\end{figure}

In Fig.~\ref{cf} we have included, for the sake of comparison, 
the mean and 1 $\sigma$ errors of $\xi(r)$ for all the
realizations of each $N$--body simulation.
The estimator used is that of Rivolo (1986)
\begin{equation}
1+\xi(r) = \sum_{i=1}^{N} \frac{N_i(r)}{n V_i},
\label{xi}
\end{equation}
with $N_i(r)$ being the number of neighbours of the $i$th point in a
shell centred at that point and having radii $r-dr/2$ and $r+dr/2$, 
and $V_i$ being the volume of the intersection of that
shell with the region $D$. There exists
an analytical expression for $V_i$ when $D$ is a cube \cite{rana}.
The estimator in equation (\ref{xi}) gives smaller errors than the usual
Davis \& Peebles (1983) estimator. This question is being addressed at
length in Pons--Border\'{\i}a et al. (1998).
In any case, the errors at large distances ($ r > 10 \, h^{-1}$ Mpc) are
very important. It is precisely in that region where it is especially
interesting to use $K$, since $\xi$ is dominated by the 
noise there and, on the contrary,
errors on $K$ are acceptable up to at least 1/3 of the box sidelength.
Although $K$ is not as informative as $\xi$, in the same way that a 
distribution function is less informative than a density, it is better
to have some information than to have none.

\begin{figure}
\vspace{1.cm}
\caption{The same as Fig. 3
for the three simulations of the cluster distribution}
\label{kcroft}
\end{figure}

In Fig.~\ref{kcroft} we present the results of the $K$ function and the
local dimension $D_2$ for the cluster simulations. It is quite evident
that no clear plateau $D_2$ is observed, although
two different regimes in its behaviour are appreciated.
At small scales, ($r \le 30-40$ $h^{-1}$ Mpc), $D_2$ increases
with the scale very slowly, with some cases showing behaviour
compatible with a constant value of $D_2$ within the uncertainties. 
Nevertheless, the systematic increasing trend is observed for the three cases.
In the second regime, for $r \ge 40$ $h^{-1}$ Mpc, the tendency to
attain homogeneity is more clearly appreciated and $D_2$ increases more 
rapidly with the scale.
This might be a good probe in order to test 
a possible fractal character of the geometry of the Universe.

It is interesting to observe how the MDM model and $\Lambda$--CDM have a
similar effect with respect to the standard CDM, namely the increase of
the amount of
clustering at all scales and the delay of the achievement of
homogeneity ($D_2 \simeq 3$). The scale at which this happens is
$70 h^{-1}$ Mpc for MDM and $\Lambda$--CDM instead of $50 h^{-1}$ Mpc
for CDM. Note that the stronger the clustering is, the
larger the values of the $K$ function.
We see that, among the three $N$--body simulations,
the strongest clustering is observed
in MDM, followed by the $\Lambda$--CDM model
and finally by the standard CDM. Regarding the dimension $D_2$,
standard CDM and $\Lambda$--CDM show similar values at small scales,
while at large scales the
agreement is between the MDM and the $\Lambda$--CDM models.
$K$ is none the less not able to distinguish between MDM and
$\Lambda$--CDM. 

In order to test the stability of this statistic when applied
to smaller regions, we have subdivided one of the cubes 
(of 300 $h^{-1}$ Mpc sidelength) containing a
MDM simulation in 8 smaller cubes of 150 $h^{-1}$ Mpc sidelength.
We have calculated the $K$ function for each of the 
subcubes up to 50 $h^{-1}$ Mpc and, as we can see in Fig.~\ref{ksubcub},
the statistic and its estimator are quite stable, since the results
for the whole sample lie within the 1 $\sigma$ deviation from the
average of $K$ calculated on the smaller cubes. 
This is an illustration of the robustness of the estimator.

\begin{figure}
\vspace{1.cm}
\caption{The average and standard deviations of the $K$ function for
8 subcubes of side 150
$h^{-1}$ Mpc extracted from one of the MDM realizations.
The dotted line corresponds to the whole simulation.}
\label{ksubcub}
\end{figure}

\section{The $K$ function on galaxy redshift surveys}
\subsection{The surveys}
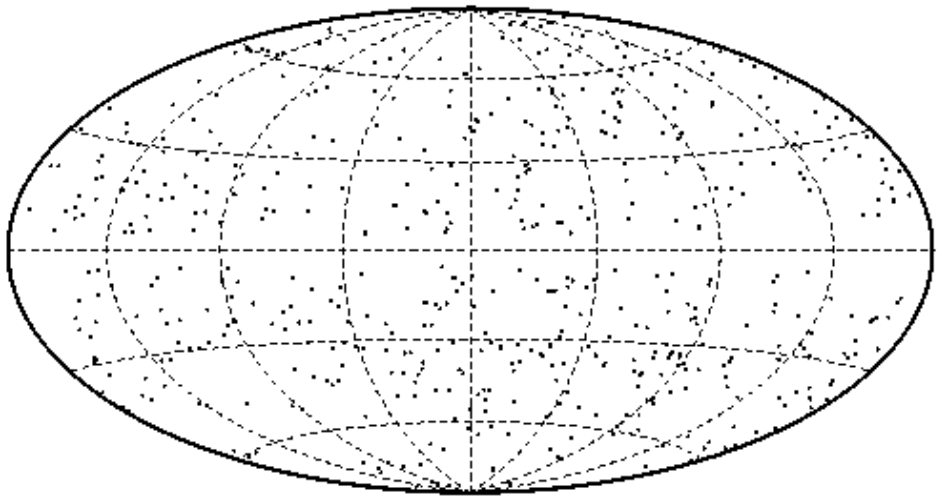
\begin{figure}
\vspace{1.cm}
\caption{Equal--area projections
of some of the volume--limited samples analysed here:
a) PPS (the north celestial
pole is at the centre of the plot); b) The volume--limited sample of
the Stromlo--APM redshift survey (now the south galactic
pole is at the centre of the plot); c) Aitoff projection of
the volume--limited sample of the {\it IRAS} 1.2 Jy survey (in
galactic coordinates).}
\label{galax}
\end{figure}

In order to check if the $K$ function produces reliable results
when used on real data sets, we have applied it to four
different galaxy catalogues.
In all cases we have extracted volume--limited samples
which are shown in Fig.~\ref{galax}; redshifts 
had been corrected from the heliocentric velocity with respect to the 
microwave background. Let us briefly describe the analysed samples:

\subsubsection{CfA-I}

The CfA-I catalogue with magnitude limit $m_B=14.5$,
compiled by Huchra et al. (1983) was based on Zwicky's angular catalog.
We have extracted a complete volume--limited sample,
which we shall call CfA80, lying inside the region
delimited by galactic latitude $b>40^\circ$, declination
$\delta>0$, distance to the Earth greater
than 17\h \ Mpc and less than 80\h\ Mpc. It contains 185
galaxies, being the Coma cluster the
most prominent feature in the sample. The volume comprised by it is
$\simeq 3.13\times 10^5 (h^{-1}$\ Mpc$)^3$ and subtends a solid angle of
$\omega=1.832605$ sr.

\subsubsection{Perseus--Pisces Supercluster (PPS)}

The Perseus--Pisces catalogue has been compiled by Giovanelli \&
Haynes (1991),
taking as a basis the old Zwicky catalogue. Its magnitude limit
is $m_Z=15.7$ and it covers (in equatorial coordinates) the region
$\alpha\in[22^h,24^h]\cup [0^h,4^h]$, $\delta\in
[0^{\circ},50^{\circ}]$. 
The most important feature contained in it is the Perseus--Pisces filamentary 
supercluster. The catalogue magnitudes have been corrected from
interstellar absorption.  

The extraction from this catalogue of a volume--limited sample has been
performed by several authors, (see Ghigna et al. (1994) and references 
therein). We use the volume--limited sample extracted by Kerscher et al.
(1997).
They have neglected the zone most affected by
galactic obscuration and restricted the sample to the area
$\alpha\in[22^h.5,24^h]\cup[0^h,3^h]$, $\delta\in[0^\circ,40^\circ]$.
The depth is 79\h\ Mpc, it contains 817 galaxies and covers a solid
angle of 0.76 sr, being the volume $\simeq 1.24 \times 10^5$ ($h^{-1}$ Mpc)$^3$.

\subsubsection{Stromlo--APM}

The Stromlo--APM redshift survey was compiled by Loveday et al. (1996),
based on the APM Bright Galaxy Catalogue \cite{lov}. It consists of 1797
galaxies with a magnitude limit of $b_{J} \le 17.15$ selected randomly
at a rate of 1 in 20. The survey covers 4300 square degrees of the
southern galactic sky, approximately defined in equatorial coordinates by 
$\alpha \in [21^{h},24^{h}] \cup [0^{h},5^{h}]$, $\delta \in
[-72.5^{\circ},-17.5^{\circ}]$. The survey redshifts have been
transformed to the Local Group reference frame and $K$--corrections have
been applied for different morphological types in the $b_{J}$ system.

We have extracted a sample volume--limited to $200h^{-1}$ Mpc (assuming
$q_{0}=0.5$), consisting of 387 galaxies. 
Distances are calculated according to the Mattig formula 
which for this choice reads as:
\begin{equation}
r= {2c \over H_{0}} \left ( 1 - {1 \over \sqrt{1+z}} \right ).
\end{equation}

The fact that the sample is not complete is not a problem for the
calculation of $K$ since the function $K$ is invariant under
thinning (van Lieshout \& Baddeley 1996).

\subsubsection{IRAS 1.2 Jy}

The {\it IRAS} 1.2 Jy survey was compiled by Fisher et al. (1995) based on the
{\it IRAS} Point Source Catalogue (PSC) by Beichman et al. (1985). It contains 
5321 galaxies and is complete
to a flux limit of 1.2 Jy at 60$\mu$m. The survey covers 87.6\% of the
sky, excluding only the Galactic plane region $\mid b \mid < 5^{\circ}$,
a small region of sky not surveyed by IRAS and confused regions in the
PSC. 

We have extracted a sample volume--limited to 120$h^{-1}$ Mpc from the
survey. It contains 561 galaxies, comprising 270 galaxies in the
northern galactic hemisphere and 291 galaxies in the southern hemisphere.

\subsection{Results}
We have typically analysed the $K$ function up to 1/2 of the cubic
root of the volume of each sample.
The weighing term $\omega_{ij}$ in the denominator of the estimator $\hat
K_R$ in equation (\ref{estim}), which depends upon the geometry, 
is now measured
by Monte Carlo integration since the shapes of the galaxy samples are
not simple cubes but something much more complicated.

\subsubsection{Optical samples}

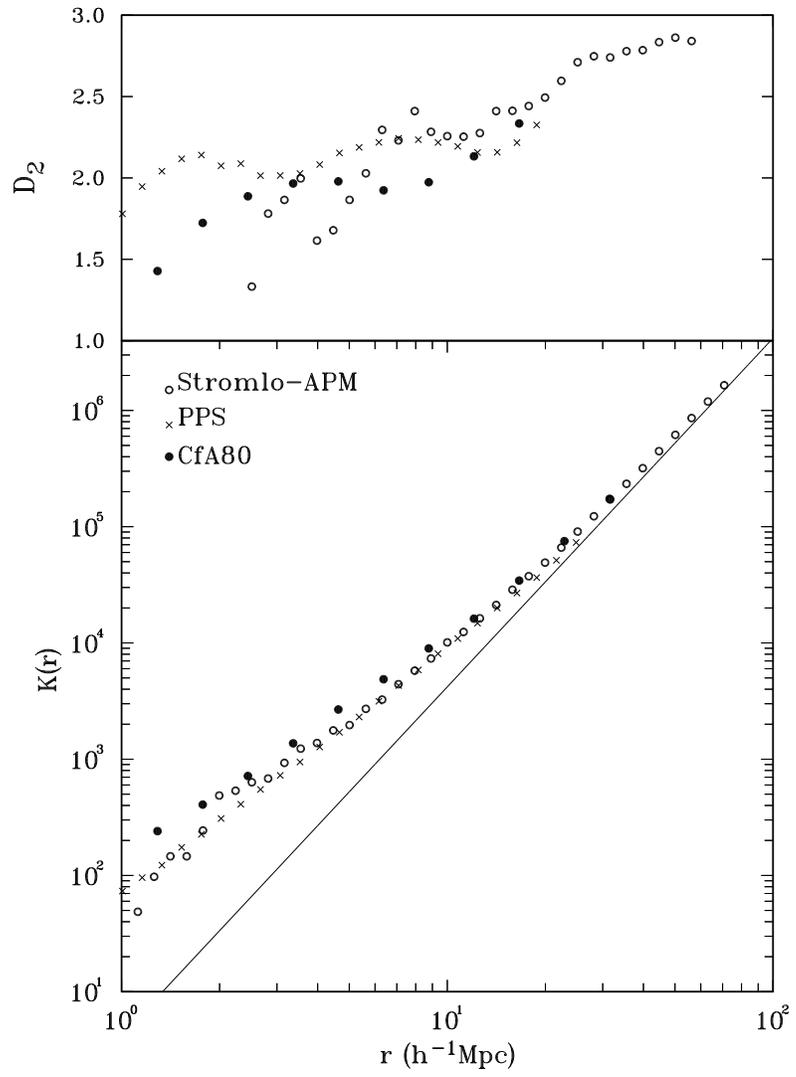
\begin{figure}
\vspace{1.cm}
\caption{The $K$ function for the CfA80 sample, the 
PPS sample and the Stromlo--APM survey. On top we show the
local dimension $D_2$ (only where the correlation coefficient was
at least 0.97) as a function of the scale $r$.}
\label{koptic}
\end{figure}

In Fig.~\ref{koptic} a plot of the results for the optically selected 
samples can be seen. On the top panel the results for the correlation dimension 
$D_2$ are shown, calculated where the function $K$ fits reasonably well a
power--law.  

In the CfA80 sample, due to its shallowness,
we have calculated $K$ only up to
$\simeq 30$\h\ Mpc but it is enough to witness its transition to homogeneity.
It is also remarkable the rapid change of its $D_2$, which increases
from 1.3 to almost 2.5 in less than 10\h\ Mpc.

The Perseus--Pisces survey presents clustering at all the scales we could
analyse but it tends to homogeneity with increasing $r$, since
its $K$ value tends to $K_{\rm \tiny Pois}$.
The PPS sample is contained in too a small region
to allow us inspection of $K$ at very large scales. We have
calculated the $K$ function just up to 25 $h^{-1}$ Mpc. 
Although it could be possible to fit
a single power--law to the behaviour of the $K$ function over the
whole range with reasonable accuracy (Guzzo et al. 1991; Bonometto et
al. 1994; Pietronero, Montuori \& Sylos--Labini 1996), we are able
to detect a decrease in the amount of clustering through
the use of the local dimension $D_2$, which goes from 1.8 at scales around 
1 $h^{-1}$ Mpc to 2.3 at scales around 20 $h^{-1}$ Mpc.
In order to go beyond 20 $h^{-1}$ Mpc we have to analyse the other
deeper samples.

At small scales the values of $K$ for the Stromlo--APM sample
are noisier due to the sparseness of the sample.
In the range [5--25] $h^{-1}$ Mpc, however, where the estimates are more 
reliable, there is  a reasonable agreement among the results for the three 
samples.

It is interesting to notice that there is a kind of continuity between the 
$D_2$ values for the three optical samples. One can form an increasing curve 
which begins at small scales sampled by CfA80 with a value $D_2\simeq 1.3$, 
joins the value $D_2\simeq 2.2$ of PPS at intermediate scales
and finally approaches the homogeneity value $D_2\simeq 3$ with APM (which
is the deepest sample), ruling out the idea of an unbounded fractal universe.
This result is in agreement with a recent paper by Scaramella el at. (1998)
in which they found evidences for a $D_2 = 3$ dimensionality for the
ESO Slice project redshift survey and the Abell (ACO) clusters, using,
as we do, comovil cosmological distances and $K$--corrections.
Similar conclusions are reached by Wu, Lahav \& Rees (1998) from the analysis
of the X-ray Background and the Cosmic Microwave Background.

The increasing value of $D_2$ with the scale is less
evident, although appreciable, for the PPS sample, in which the
behaviour is a smooth increasing trend with oscillations
that might be misinterpreted as a constant. In any case,
it is interesting to remark that this survey has been
selected to isolate one strong feature in the local Universe
(Iovino et al. 1993), the Perseus--Pisces supercluster, a
very big sheet--like structure, which contributes strongly
to values close to $2$ for the correlation dimension. However, as we
have seen with the other analysed samples, this behaviour is particular
of this sample due to a selection effect, and cannot be extrapolated
to the whole Universe.

\subsubsection{IRAS 1.2 Jy}
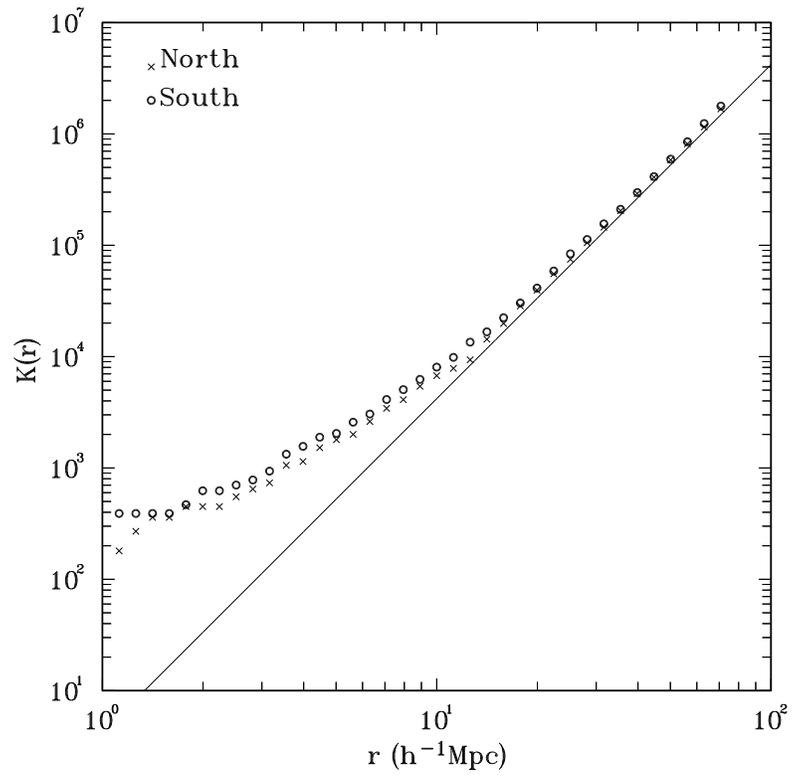
\begin{figure}
\vspace{1.cm}
\caption{The $K$ function for the volume--limited samples of the
1.2 Jy {\it IRAS} catalogue. We use different marks for the North and
South Galactic hemispheres. The differences between both are significant at
small scales.}
\label{kiras}
\end{figure}

Here we have analysed separately both galactic hemispheres
by means of the $K$ function. In Fig.~\ref{kiras}
we see significant differences
between the $K$ function of the North and the South samples
up to 20 $h^{-1}$. The clustering seems stronger in the Southern
hemisphere than in the Northern one. This result corroborates the findings
of Kerscher et al. (1997). These authors found differences in the
strength of the clustering between both hemispheres by means of the Minkowski
functional and nearest--neighbour distributions.
Also at small scales the values of $K$ are typically
larger for the {\it IRAS} 1.2 Jy sample than for the optical
samples analysed previously. This result is in full agreement with
the estimates of the correlation function at small scales reported in
Fisher et al. (1994), Bonometto et al. (1994) and Loveday et al. (1995).
However, at larger scales the $K$ functions of
both hemispheres approach in the same way the curve corresponding to
a uniform Poissonian distribution.

\subsection{Reliability of the results}

One way of measuring the errors on the estimates of a statistical
function such as $K(r)$ is based on the dispersion of its measures
when applied to ensembles of artificial catalogues having
similar statistical properties (Fisher et al. 1993; Hamilton 1993).
The segment Cox processes \cite{skm} are quite useful in this context.
The particular model of Cox process (see also Pons--Border\'{\i}a et al. 1998)
used here has been generated in the following way: we scatter randomly in the
space segments of length $l$,
with $\lambda_s$ being the mean number of segments per unit volume and, on
those segments, we randomly distribute points so that a chosen
$\lambda_l$ be the mean number of points per unit length of the segments. 
The intensity of this process 
will be $\lambda=\lambda_l \lambda_s l$ and, as proven in Stoyan, Kendall \&
Mecke (1995), its $K$ function will be given by:

\begin{equation}
K(r)= \left\{\begin{array}{ll}
        \frac{4 \pi}{3} r^3 + \frac{r}{\lambda_s l} \left ( 2 - {r \over l}
        \right ) & \mbox{if $r \le l$} \\
        \frac{4 \pi}{3} r^3 + \frac{1}{\lambda_s} & \mbox{if $r > l$} 
        \end{array} \right.
\label{kcox}
\end{equation}
and one will be able to calculate $D_2$ analytically simply as:

\begin{equation}
D_2(r)=\frac{r}{K(r)} \times \frac{dK(r)}{dr} .
\label{d2cox}
\end{equation}

We have generated 10 such Cox processes containing between 1400 and 1600 
points each inside a cube of side 100 $h^{-1}$ Mpc. The values we have
taken for the parameters are $l=20, \lambda_s=4\times 10^{-5}, \lambda_l=1.88, 
\lambda\simeq 1500/100^3$. 

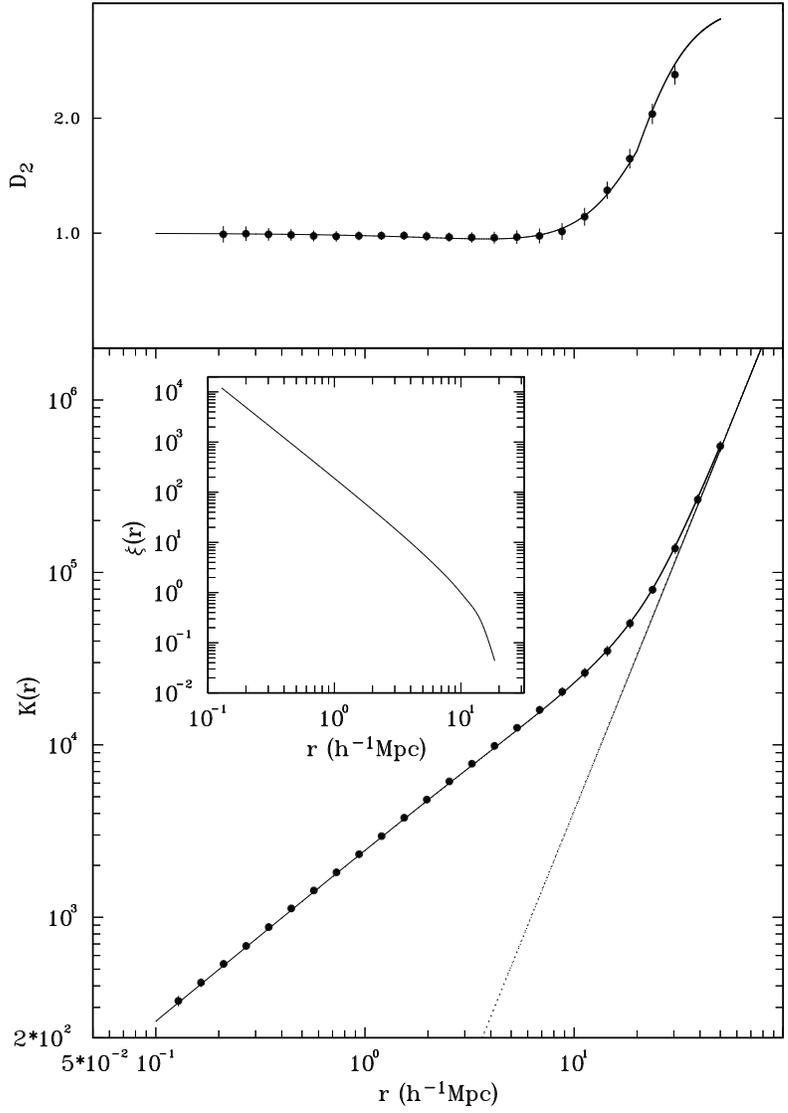
\begin{figure}
\vspace{1.cm}
\caption{Bottom panel: the average and 1 $\sigma$ deviation error bars
of the function 
$K(r)$ for 10 realizations of the 
Cox point processes (solid discs).
The inset shows the two--point correlation
function of this stochastic model. Top panel: the local
correlation dimension $D_2$ with 1 $\sigma$ uncertainties
calculated by means  of a five--point weighted log--log least square
fit on the average of $K$.
In both panels the solid line shows the theoretical values, while the dotted 
line in the bottom panel corresponds to $K_{\rm \tiny Pois}(r)$.
} 
\label{kcoxfig}
\end{figure}

In Fig.~\ref{kcoxfig} we can see the average empirical values of the
estimates
of $K(r)$ obtained by means of equation (\ref{estim}) together with
the standard deviations. The expected theoretical function given by
equation (\ref{kcox}) is depicted as a solid line.
As we can see, an empirical estimate of $K$ calculated on 10 Cox processes
reproduces quite satisfactorily the expected theoretical behaviour. 
Note that the variance of the number of counts in a bounded set
for a Cox process is always bigger than in 
a Poisson process having the same intensity \cite{skm}.
It is important to notice that the border correction has not destroyed 
the goodness of the estimator; in particular, it has not introduced
spurious homogeneization. Our estimator has worked
successfully not only in the ``easy'' case of absence of structure 
which represent Poisson processes but
it has also been able to reproduce quite exactly the very precise
value of $K$ for a clustered Cox process. This test gave us enough
confidence to believe that the $K$ results obtained from the galaxy samples
effectively reflect the structure existing there.

The Cox processes used here also play an interesting role.
They are a good example with which to prevent the naive use of fractals in
the analysis of point fields as it has been anticipated by Stoyan (1994).
In the inset of Fig.~\ref{kcoxfig}, we can see the function $\xi(r)$
expected for this process. The function is the sum of two power--laws
$\xi(r)=Ar^{-2}+Br^{-1}$ with $A=(2\pi\lambda_s l)^{-1}$ and
$B=-(2\pi\lambda_s l^2)^{-1}$. At short distances the first power--law
dominates and therefore the function $\xi(r)$ can be nicely fitted
to a power--law $\xi(r) \propto r^{-\gamma}$ with $\gamma=2$
and this could lead to
interpret that the point set is a fractal when clearly it is not
(Stoyan 1994; McCauley 1997).
Because in this regime $\xi(r) \gg 1$, the same can be said for
the function $1+\xi(r)$. Looking at the top panel of Fig.~\ref{kcoxfig},
we can see the behaviour of the empirical local dimension $D_2$
calculated  over the average of the 10 realizations of the Cox processes
together with the 1 $\sigma$ deviations. The solid line represents the
expected theoretical values [equation (\ref{d2cox})].
Again we can see the reliability of the
estimates. But what it is more interesting in this example is
the long plateau observed in the plot of the correlation dimension. The
value $D_2 \simeq 3 -\gamma \simeq 1$ remains nearly constant for a broad
range of scales, due to the particular
behaviour of the $K$ function for this model. After the ``fractal"
behaviour, a transition to homogeneity is clearly appreciated in both
$D_2$ and $K(r)$. It is interesting to remark the qualitatively
similar behaviour between this figure and Fig.~\ref{koptic} in which
we showed the same function for the analysed optical redshift
surveys: a regime at small scales
where the clustering is strong (with $K \gg K_{\rm \tiny Pois}$) and
where $K$ can be fitted to a power--law. At larger scales, however 
the increasing behaviour of the local dimension $D_2$ with the scale and the
continuous approximation of the function $K$ to $K_{\rm \tiny Pois}$ are 
absolutely clear.  At this point we want to remark that, in the same
way that the term fractal is not appropriate for the Cox process, even
having a correlation function decaying as a power--law at short scales
\cite{stofrac}, the galaxy distribution, even holding a similar
property, is not a fractal in a rigorous sense \cite{mccau}. However in a
more loose use of the term fractal \cite{avn98}, 
it could be appropriate to talk about a ``fractal" regime to 
describe the range of scales where $K(r)$ follows
a power--law, bearing in mind that a real self--similar point pattern,
for example the Soneira--Peebles model described in section 3,
verifies other conditions (self--similarity) apart from a power--law 
decaying correlation function.

\section{Conclusions}

We should like also to comment briefly on the relation of $K$ with the
correlation function $\xi(r)$. 
Both play their role in the analysis of the point pattern and, 
as Stoyan \& Stoyan (1996) say, their relation is
similar to that between the distribution function and the
probability density function in classical statistics. The use
of a cumulative quantity such as $K$ avoids binning in
distance, which is often a source of arbitrariness for $\xi$ \cite{rip92}.  
Let us explain why $\xi$ does suffer from the hindrance of splitting the
information into disjoint bins.
When one estimates $\xi(r)$ in $[r,r+dr]$, it is assumed that
within that bin the correlation function is constant, and since this is 
obviously not true, the larger the bin the larger the error, but 
we cannot make arbitrarily small the size $dr$ of the bin,
because in that case we would not find any pairs.
In other words, $\xi(r)$ has an additional source of bias, not present
in $K$, due to the smoothing caused by averaging over pairs of points
close to but not exactly $r$ units apart of each other (Stein 1996).

The correlation length ($r_0| \xi(r_0)=1$) is 
just the scale at which the density of 
galaxies is, on average, twice the mean number density. At smaller scales 
the pair correlations are due to non--linear perturbations, but homogeneity 
is not reached till $\xi(r_{\rm \tiny hom}) \sim 0$. 
The main interest of $K$ is that it permits us to study clustering 
precisely in that \lq\lq difficult'' 
range where $r_0 < r < r_{\rm \tiny hom}$,
which cannot be reached by $\xi$ because in this range the errors on the 
estimates of $\xi$ are comparable with their 
values, while the difference $K-K_{\rm \tiny Pois}$ is still meaningful. 

As a concluding remark, we want to stress that an unbiased estimator of
a quantity related with the correlation integral, known as the
$K$ function, has been applied to cosmological simulations and galaxy
samples. This function, extensively used in the field of spatial statistics,
provides a nice measure of clustering. The border correction used
here does not waste any data points and does not introduce spurious
homogeneization,
giving reliability to the evaluation of this function at large scales.
Through the slope of $K$ we are able to calculate $D_2$, which is
an indicator of a possible fractal behaviour
of the point process at a given scale range.
The clear physical meaning of $K$ and $D_2$
helps us easily interpret the clustering properties of different
models of structure formation at different scales.

Regarding the analysis of the galaxy redshift surveys, we have seen that
the estimator of the $K$ function is robust in the sense that it does
not depend on the shape of the study region and provides us with reliable
information about the point patterns over a wide range of scales. The
behaviour of the local dimension $D_2$ for 
the real galaxy samples is particularly interesting 
to proponents of various fractal models of large--scale structure.
If a constancy of $D_2$ with the scale is a necessary condition for having a
fractal point pattern (although it should not be sufficient as we have seen
with the Cox process [see also Stoyan (1994) for more examples]), it
is a neat conclusion of our analysis that the galaxy distribution does
not even hold the necessary condition.
The analysis presented here will provide a conclusive
test to discover the scale at which the distribution of the matter
in the Universe is really homogeneous when applied, in the near future,
to the bigger and deeper galaxy catalogues which will be soon ready for
common use.

\subsection*{ACKNOWLEDGEMENTS}

This work has been partially supported by an EC Human Capital and
Mobility network (contract ERB CHRX-CT93-0129) and by the Spanish
DGES project n. PB96-0707.
We thank prof. Stoyan for bringing the Cox model to our attention
and for useful conversations and comments.
We thank R. Croft, S. Paredes, R. Trasarti--Battistoni  and R. 
van de Weygaert  for kindly allowing us to use their samples and programs, 
as well as T. Buchert, J. Schmalzing, M. Stein and specially M. Kerscher 
for very interesting discussions and comments.
The authors want to thank the anonymous referee for his/her valuable 
comments and suggestions.


\begin{thebibliography}{99}

\bibitem[Avnir et al. 1998]{avn98} Avnir, D., Biham, O., Lidar, D.,
\& Malcai, O. 1998, Science 279, 39
\bibitem[Baddeley et al. 1993]{rana}  Baddeley, A.J., Moyeed,
R.A., Howard, C.V., \& Boyde, A. 1993, Appl. Statist. 42, 641
\bibitem[Bartlett 1964]{bar64} Bartlett, M.S. 1964, Biometrika 51, 299
\bibitem[Beichman et al 1985]{bei95} Beichman, C.A., Neugebauer, G., 
Habing, H.J., Clegg, P.E., \& Chester, T.J., 1985, 
IRAS Catalogs and Atlases, Washington
\bibitem[Bonometto et al 1994]{bono94} Bonometto, S., Iovino, A.,
Guzzo, L., Giovanelli, R. \& Haynes, M., 1994, ApJ, 419, 451
\bibitem[Borgani et al. 1996]{bor96} Borgani, S., Mart\'{\i}nez, V.J., 
P\'erez, M.A., \& Valdarnini, R. 1994, ApJ 435, 37
\bibitem[Buchert \& Mart\'{\i}nez 1993]{buc93} Buchert, T., \& Mart\'{\i}nez,
V.J. 1993, ApJ 411, 485
\bibitem[Cappi et al. 1998]{cap98} Cappi, A., Benoist, C., da Costa, L.N., \&
Maurogordato, S. 1998, astro--ph/9804085, A\&A (in press)
\bibitem[Coleman \& Pietronero 1992]{col92}  Coleman, P.H.,
\& Pietronero, L. 1992, Phys. Rep., 213, 31
\bibitem[Croft \& Efstathiou 1994a]{cr94a}  Croft, R.A.C.,
\& Efstathiou, G. 1994a, MNRAS 267, 390
\bibitem[Croft \& Efstathiou 1994b]{cr94b} Croft, R.A.C.,
\& Efstathiou, G. 1994b, MNRAS 268, L23
\bibitem[Davis \& Peebles 1983]{dp83} Davis, M., \& Peebles, P.J.E.
1983, ApJ 267, 465
\bibitem[Diggle 1983]{dig83}  Diggle, P.J., 1983, Statistical Analysis
of Spatial Point Patterns. Academic Press, London
\bibitem[Doguwa \& Upton 1989]{dog89} Doguwa, S.I. \& Upton, G.J.G.. 1989,
Biom. J. 5, 563 
\bibitem[Dom\'{\i}nguez--Tenreiro, G\'omez--Flechoso \& Mart\'{\i}nez 1994]
{dom94} Dom\'{\i}nguez--Tenreiro, R., G\'omez--Flechoso, M.A. \& 
Mart\'{\i}nez, V.J. 1994, ApJ 424, 42
\bibitem[Dubuc et al. 1989]{dub89} Dubuc, B., Quiniou, J.F, Roques--Carmes,
C., Tricot, C. \& Zucker, S.W. 1989, Phys. Rev A, 39, 1500
\bibitem[Einasto \& Gramann 1993]{ein93} Einasto, J., \& Gramann, M. 1993,
ApJ 407, 443
\bibitem[Fisher et al 1993]{fis93} Fisher K.B., Davis M., Strauss M.A., 
Yahil A., Huchra J.P., 1993, ApJ, 402, 42
\bibitem[Fisher et al 1994]{fis94} Fisher K.B., Davis M., Strauss M.A., 
Yahil A., Huchra J.P., 1994, MNRAS, 266, 50
\bibitem[Fisher et al. 1995]{irasref} Fisher, K.B., Huchra, J.P., 
Strauss, M.A., Davis, M., Yahil, A., \& Schlegel, D. 1995, ApJS 100, 69
\bibitem[Geller \& Huchra 1989]{gel89} Geller, M.J., \&
Huchra, J.P. 1989, Science, 246, 897
\bibitem[Giovanelli \& Haynes 1991]{ppsref} Giovanelli, R., \& Haynes,
M. 1991, ARA\&A, 29, 499
\bibitem[Ghigna et al. 1994]{rob} Ghigna, S., Borgani, S., Bonometto,
S.A., et al. 1994, ApJ Let., 437, L71
\bibitem[Guzzo et al 1991]{guz91} Guzzo, L., Iovino, A., Chincarini, G.,
Giovanelli, R. \& Haynes, M.P. 1991, ApJ, 382, L5
\bibitem[Guzzo 1997]{guz97} Guzzo, L. 1997, New Astronomy 2, 517 
\bibitem[Hamilton 1993]{ham93} Hamilton, A.J.S. 1993, ApJ, 417, 19
\bibitem[Huchra et al. 1993]{huc83} Huchra, J.P., Davis, M., 
Latham, D., \& Tonry, J. 1983, ApJ Suppl., 52, 89
\bibitem[Iovino et al. 1993]{io93} Iovino, A., Giovanelli, R., Haynes, M.P., 
Chincharini, G. \& Guzzo, L. 1993, MNRAS, 265, 21
\bibitem[Kerscher et al]{aiha} Kerscher, M., Pons--Border\'{\i}a, M.J., 
Schmalzing, J., Trasarti--Battistoni, R., Buchert, T., \&
Mart\'{\i}nez, V.J. 1997, astro-ph/9712098
\bibitem[Kerscher et al]{ker98} Kerscher, M., Schmalzing, J., Buchert, T.,
\& Wagner, H. 1998, A\&A 333, 1 
\bibitem[Kirshner et al. 1981]{kir81} Kirshner, R.P., Oemler, A.
Jr., Schechter, P.L., \& Shectman, S.A. 1981, ApJ 248, L57
\bibitem[Loveday et al. 1995]{lov95} Loveday, J., Maddox, S.J.,
Efstathiou, G., \& Peterson, B.A. 1995, ApJ 442, 457
\bibitem[Loveday et al. 1996]{lov96} Loveday, J., Peterson, B.A., 
Maddox, S.J., \& Efstathiou, G. 1996, ApJ Suppl 107
\bibitem[Loveday 1996]{lov} Loveday, J. 1996, MNRAS 278, 1025
\bibitem[Mart\'{\i}nez et al. 1990]{mar90}  Mart\'{\i}nez, V.J.,
Jones, B.J.T., Dom\'{\i}nguez-Tenreiro, R., \& van de Weygaert, R. 1990,
ApJ 357, 50
\bibitem[Mart\'{\i}nez \& Coles 1994]{mar94} Mart\'{\i}nez, V.J., \&
Coles, P. 1994, ApJ 437, 550
\bibitem[Mart\'{\i}nez et al. 1995]{sci} Mart\'{\i}nez, V.J.,
Paredes, S., Borgani, S., \& Coles, P. 1995, Science 269, 1245
\bibitem[McCauley 1997]{mccau} McCauley, J.L. 1997, astro/ph 9703046
\bibitem[Neymann \& Scott 1952]{ney52} Neymann, J., \& Scott,
E. 1952, ApJ 116, 144
\bibitem[Neymann \& Scott 1955]{ney55} Neymann, J., \& Scott,
E. 1955, AJ 60, 33
\bibitem[Park et al. 1994]{par94} Park, C., Vogeley, M.S., Geller, M.J.,
\& Huchra, J.P. 1994, ApJ 431, 569
\bibitem[Peebles 1980]{p80}  Peebles P.J.E. 1980,
The Large Scale Structure of the Universe. Princeton University
Press, Princeton
\bibitem[Peebles 1989]{peb89}  Peebles, P.J.E. 1989, Physica D,
38, 273
\bibitem[Peebles 1993]{p93} Peebles, P.J.E. 1993,
Physical Cosmology. Princeton University Press, Princeton
\bibitem[Pietronero et al. 1996]{pie96}
Pietronero, L., Montuori, M., \& Sylos Labini, F., 1996, in Turok N., ed.,
Critical Dialogues in Cosmology.
to appear, astro-ph/9611197
\bibitem[Pons--Border\'{\i}a et al. 1998]{nos} Pons--Border\'{\i}a, M.J.,
Mart\'{\i}nez, V.J., Stoyan, D., Stoyan, H., \& Saar, E.,
1997, submitted
\bibitem[Rivolo 1986]{riv86} Rivolo, A.R. 1986, ApJ, 301, 70
\bibitem[Ripley 1976]{rip76}  Ripley, B.D. 1976,
J. Appl. Prob., 13, 255
\bibitem[Ripley 1977]{rip77}  Ripley, B.D. 1977,
J. Royal Statist. Soc., 39, 172
\bibitem[Ripley 1981]{rip81}  Ripley, B.D., 1981, Spatial Statistics.
John Wiley \& Sons, NY
\bibitem[Ripley 1992]{rip92}  Ripley, B.D. 1992, in
Feigelson, E.D. \& Babu, G.J., ed.,
Statistical Challenges in Modern Astronomy. Springer--Verlag, NY, p. 102
\bibitem[Scaramella et al. 1998]{sca98} Scaramella, R., et al. 1998, 
A\&A, in press (astro/ph 9803022)
\bibitem[Smoot et al. 1992]{cobe}  Smoot, G.F., et al. 1992, ApJ,
396, L1
\bibitem[Soneira \& Peebles 1978]{sp78} Soneira R.M., \& Peebles
P.J.E. 1978, AJ 83, 845
\bibitem[Stein 1996]{ste96}  Stein, M. 1996,
in Feigelson, E.D. \& Babu, G.J., ed.,
Statistical Challenges in Modern Astronomy II. Springer--Verlag, NY
\bibitem[Stoyan, Kendall \& Mecke 1995]{skm} Stoyan, D., Kendall, W.S.,
\& Mecke, J. 1995, Statistic Geometry and its Applications.
Springer--Verlag, Berlin
\bibitem[Stoyan 1994]{stofrac} Stoyan, D. 1994, Statistics 25, 267
\bibitem[Stoyan \& Stoyan 1994]{sto94} Stoyan, D., \& Stoyan, H., 1994,
Fractals, Random Shapes and Point Fields. John Wiley \& Sons, Chichester
\bibitem[Stoyan \& Stoyan 1996]{sto96} Stoyan, D., \& Stoyan, H. 1996,
Biom. J. 38, 259
\bibitem[Stoyan \& Stoyan 1998]{sto98} Stoyan, D., \& Stoyan, H. 1998,
preprint 98--3 Technische Universit\"at Bergakademie Freiberg
\bibitem[Szapudi \& Szalay 1997]{sza97} Szapudi, I., \& Szalay, A.S.
astro-ph/9704241
\bibitem[Tadros \& Efstathiou 1996]{tad96} Tadros, H., \& Efstathiou, G.
1996, MNRAS 282, 1381
\bibitem[Upton \& Fingleton 1987]{upt87} Upton, G.J.G., \& Fingleton, B.
1987, Spatial Data Analysis by Example, Vol. 1. John Wiley \& Sons, NY
\bibitem[van de Weygaert 1991]{rien}  van de Weygaert, R.,
1991, Ph.D. Thesis, Rijksuniversiteit Leiden
\bibitem[van Lieshout \& Baddeley 1996]{lie96} van Lieshout, M.N.M., \& 
Baddeley, A.J. 1996, Statistica Neerlandica, 50, 344
\bibitem[Wu et al. 1998]{wu98} Wu, K.K.S., Lahav, O., \& Rees, M.J. 1998, 
astro--ph/9804062, submitted to Nature
\end{thebibliography}
\end{document}